\documentclass[runningheads]{llncs}
\usepackage[T1]{fontenc}
\usepackage{multirow}
\usepackage{multicol}
\usepackage{graphicx}
\usepackage{adjustbox}
\usepackage[pagebackref=true,breaklinks=true,colorlinks,bookmarks=false]{hyperref}
\begin{document}
\title{Segment Anything in Medical Images and Videos: Benchmark and Deployment}
\titlerunning{SAM2 In Medical Images and Videos}
%
\author{Jun Ma\inst{1,2,5} \and Sumin Kim\inst{1}$^*$ \and Feifei Li\inst{1}$^*$ \and Mohammed Baharoon\inst{2} \and Reza Asakereh\inst{1} \and Hongwei Lyu\inst{5} \and
Bo Wang\inst{1,2,3,4,5}
}
\authorrunning{J. Ma, S. Kim, F. Li, M. Baharoon, R. Ashakereh, H. Lyu, and B. Wang}
%
\institute{Peter Munk Cardiac Centre, University Health Network, Toronto, Canada \and
Vector Institute for Artificial Intelligence, Toronto, Canada \and
Department of Computer Science, University of Toronto, Toronto, Canada \and
Department of Laboratory Medicine and Pathobiology, University of Toronto, Toronto, Canada \and
AI Hub, University Health Network, Toronto, Canada \\ 
$^*$Equal contributions
\\
}
\maketitle              
\begin{abstract}
Recent advances in segmentation foundation models have enabled accurate and efficient segmentation across a wide range of natural images and videos, but their utility to medical data remains unclear. 
In this work, we first present a comprehensive benchmarking of the Segment Anything Model 2 (SAM2) across 11 medical image modalities and videos and point out its strengths and weaknesses by comparing it to SAM1 and MedSAM. 
Then, we develop a transfer learning pipeline and demonstrate SAM2 can be quickly adapted to medical domain by fine-tuning. Furthermore, we implement SAM2 as a 3D slicer plugin and Gradio API for efficient 3D image and video segmentation. 
The code has been made publicly available at \url{https://github.com/bowang-lab/MedSAM}.
\end{abstract}


\section{Introduction}
Segmentation is one of the important fundamental tasks in medical image analysis, which is the prerequisite for many downstream tasks, such as early cancer detection~\cite{lvle-pancreasNatMed} and disease diagnosis~\cite{Yanran-HeatNatMed}. During the past decade, there has been a clear methodology trend from specialist segmentation models~\cite{UNet-Nature,shelhamer2017FCN-PAMI} to generalist segmentation models or segmentation foundation models~\cite{SAM1}\cite{2023-SEEM}. 
Specialist models are usually developed for specific anatomical structures~\cite{FLARE21-MIA,Ma-2020-abdomenCT-1K}, diseases~\cite{ma2021-COVID-Data,KiTS2021MIA,LiTS} or medical imaging modalities~\cite{ronneberger2015UNet2D,bakas2018brats,nnunet21}, which can only segment the data from the same domain as the training set. 
In contrast, foundation models are trained on large-scale and diverse datasets, which have strong capabilities to segment a wide range of objects, image types and scenarios, and even zero-shot generalization ability on images from unseen domains~\cite{FM-Stanford,FM-Shaoting}. 

Segment Anything Model (SAM)~\cite{SAM1} is the first foundation model in image segmentation, which was trained on 256 GPUs with 1.1 billion mask annotations from 11 million images. Although SAM supports commonly used interaction prompts (e.g., points, bounding boxes, and masks) for flexible segmentation and adoption across a broad spectrum of images and downstream tasks~\cite{GroundSAM}, it failed to segment medical images~\cite{SAM1-Eval-Maciej}\cite{SAM1-Eval-XinYang} because the training data (SAM-1B) only contained natural images. Medical images have a large domain gap to natural images, which have unique challenges, such as low resolution and limited image quality~\cite{tajbakhsh2020embracing-MIA}. 
Nevertheless, SAM can be easily extended to the medical domain by transfer learning~\cite{MedSAM}.

MedSAM~\cite{MedSAM} has demonstrated that fine-tuning SAM on large-scale medical images can achieve superior performance for 2D medical image segmentation, but its ability in segmenting 3D medical images and videos remains limited because it is inefficient to produce prompts for each image via an image-by-image segmentation pipeline in real practice. 
Many recent studies have explored extending SAM to volumetric image segmentation, such as adding 3D adapters~\cite{SAM3D-Wenao,SAM3D-cheng} to the image encoder, directly augmenting image encoder~\cite{SAM3D-junjun,SegVol-Fanbai} to 3D, Mixture of Experts (MoE) with gating network to select task-specific finetuned models~\cite{SAM3D-Guotai}, and text-guided 3D segmentation~\cite{Yao-sam-text}.

\begin{figure}[htbp]
\centering
\includegraphics[scale=0.18]{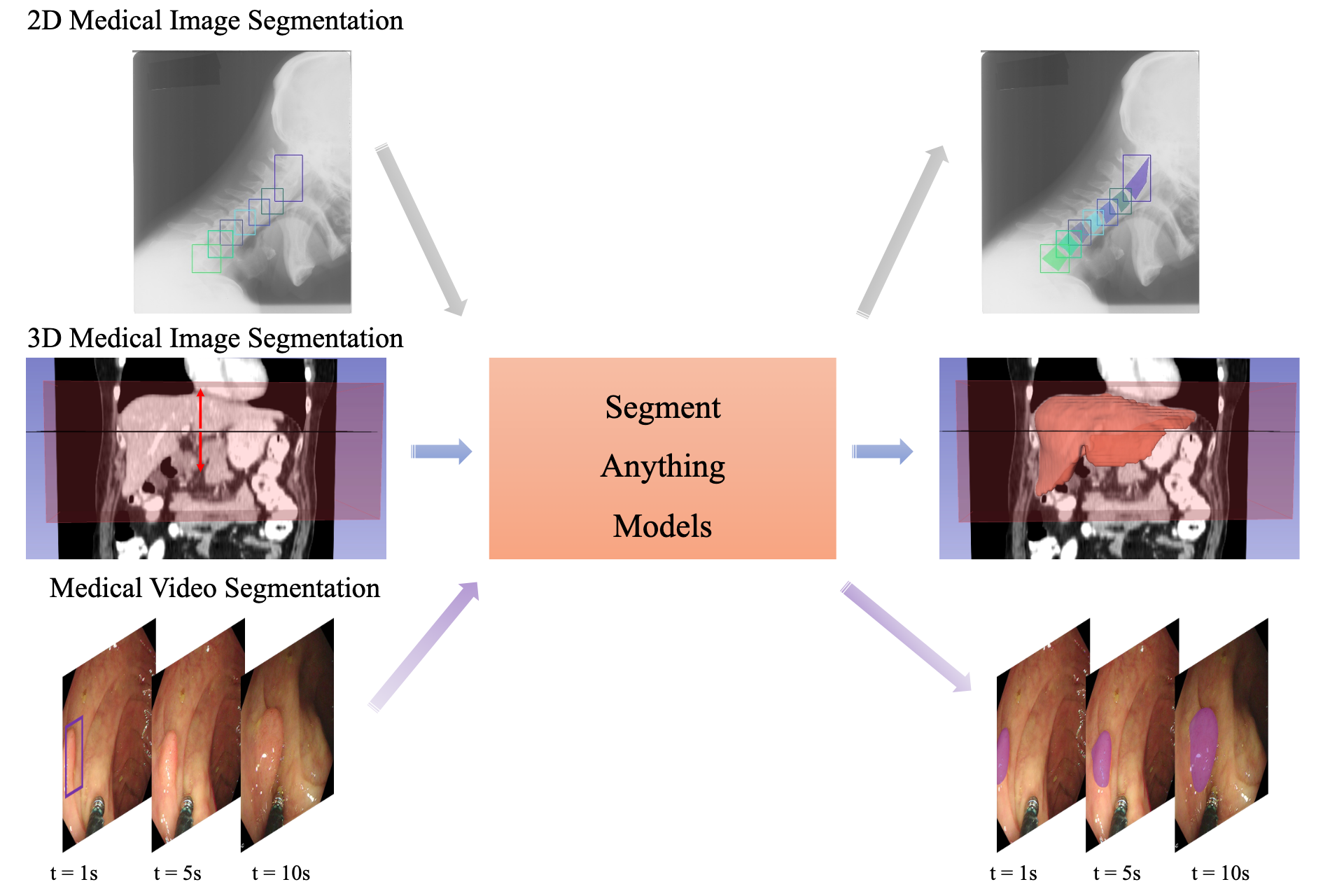}
\caption{Datasets and evaluation protocol. We evaluate SAM2 on various 2D\&3D medical images and videos. The 2D images and the bounding box prompts are directly passed to SAM2 to generate segmentation results. The 3D images and video are initialized with a bounding box prompt on the middle slice and the first frame to get the 2D masks, respectively. Then, the model propagates the 2D mask to the remaining slices/frames.}
\label{fig:pipeline}
\end{figure}

Recently, SAM2~\cite{SAM2-Eval-Maciej} has further augmented SAM to promptable video segmentation without sacrificing the image segmentation ability (Methods). The model was trained on an unprecedented dataset with 50.9K videos and outperformed existing work in established video object segmentation benchmarks. Notably, it also achieved strong zero-shot performance in many other video and image segmentation benchmarks across various distributions. However, it is not clear how SAM2 performs on medical images, especially 3D medical images and videos, since only one medical dataset (instrument segmentation in endoscopy videos~\cite{EndoVis18}) was evaluated in SAM2. Moreover, the official interface only supports evaluating short videos with limited data format, which cannot be used for medical professionals to test their own medical data with SAM2.

In this work, we address the above issues by presenting a comprehensive evaluation of SAM2 across ten medical modalities, including various 2D \& 3D images and videos. We also compare SAM2 with SAM1\footnote{We refer to the first segmentation anything model~\cite{SAM1} as SAM1, aiming to differentiate it from the recent SAM2~\cite{SAM2}.} and MedSAM to obtain a holistic understanding of their advantages and disadvantages. 
Compared to concurrent works~\cite{SAM2-Eval-Maciej,SAM2-Eval-Oxford,SAM2-Eval-Xiangfeng}, we develop a transfer learning pipeline to quickly adapt SAM2 into medical image segmentation by fine-tuning, and further incorporate SAM2 into the 3D Slicer plugin~\cite{Slicer}, allowing users to easily use SAM2 to annotate various 3D medical data (e.g., CT, MR, and PET) that are not supported in the official SAM2 interface. We also implement Gradio~\cite{Gradio} interface to support efficient medical video segmentation.

\section{Results}
\subsection{Datasets and evaluation protocol}
The benchmark dataset covers 11 commonly used medical image modalities, including computed tomography (CT), magnetic resonance imaging (MRI), positron emission tomography (PET), ultrasound (US), encoscopy, fundus, dermoscopy, mammography, light microscope, and optical coherence tomography (OCT) (Methods). 
Since SAM2 is a general model for image and video segmentation, we evaluated its ability on 2D images, 3D images, and videos. The 2D datasets contain all the modalities while the 3D datasets include CT, MR, and PET scans. The video datasets are made of US and endoscopy videos.

In addition to the recent SAM2~\cite{SAM2}, we also compare it with its predecessor SAM~\cite{SAM1} and the general medical image segmentation model (MedSAM~\cite{MedSAM}), allowing a comprehensive understanding of their performances. SAM and SAM2 have three and four different model sizes, respectively, and all of them are evaluated in the experiments. 
Similar to MedSAM, we still advocate the bounding box prompt because it is not only efficient without trial and error but also has fewer ambiguities compared to the point prompt.  
For 2D images, we pass both images and bounding box prompts to the model to get the corresponding masks.
For 3D images, we initialize the bounding box prompts in the middle slice followed by calling the models to generate the 2D masks. Since SAM and MedSAM only support 2D image segmentation, the remaining parts are segmented slice-by-slice from middle to bottom and top. The segmentation masks on the current slice are used to simulate bounding box prompts of the next slice in a sequential way.
SAM2 supports video segmentation and 3D medical images can be cast as videos where each image is one frame~\cite{MedImg-Seg3D-MedIA-Best23}. Thus, the middle slice segmentation mask is propagated to the remaining images by the video segmentation function in SAM2. We opt for the middle slice as the starting frame in 3D image segmentation because it usually has the largest object size among all the slices in the conventional axial view. 
For video segmentation, we initialize the bounding box prompt on the first slice to get the initial mask followed by passing the mask to the video segmentation model. Only SAM2 is evaluated for video segmentation because the object locations vary greatly across different frames and simply applying SAM or MedSAM in a frame-by-frame way does not work.

\subsection{Evaluation results on 2D image segmentation}
\begin{figure}[htbp]
\centering
\includegraphics[scale=0.182]{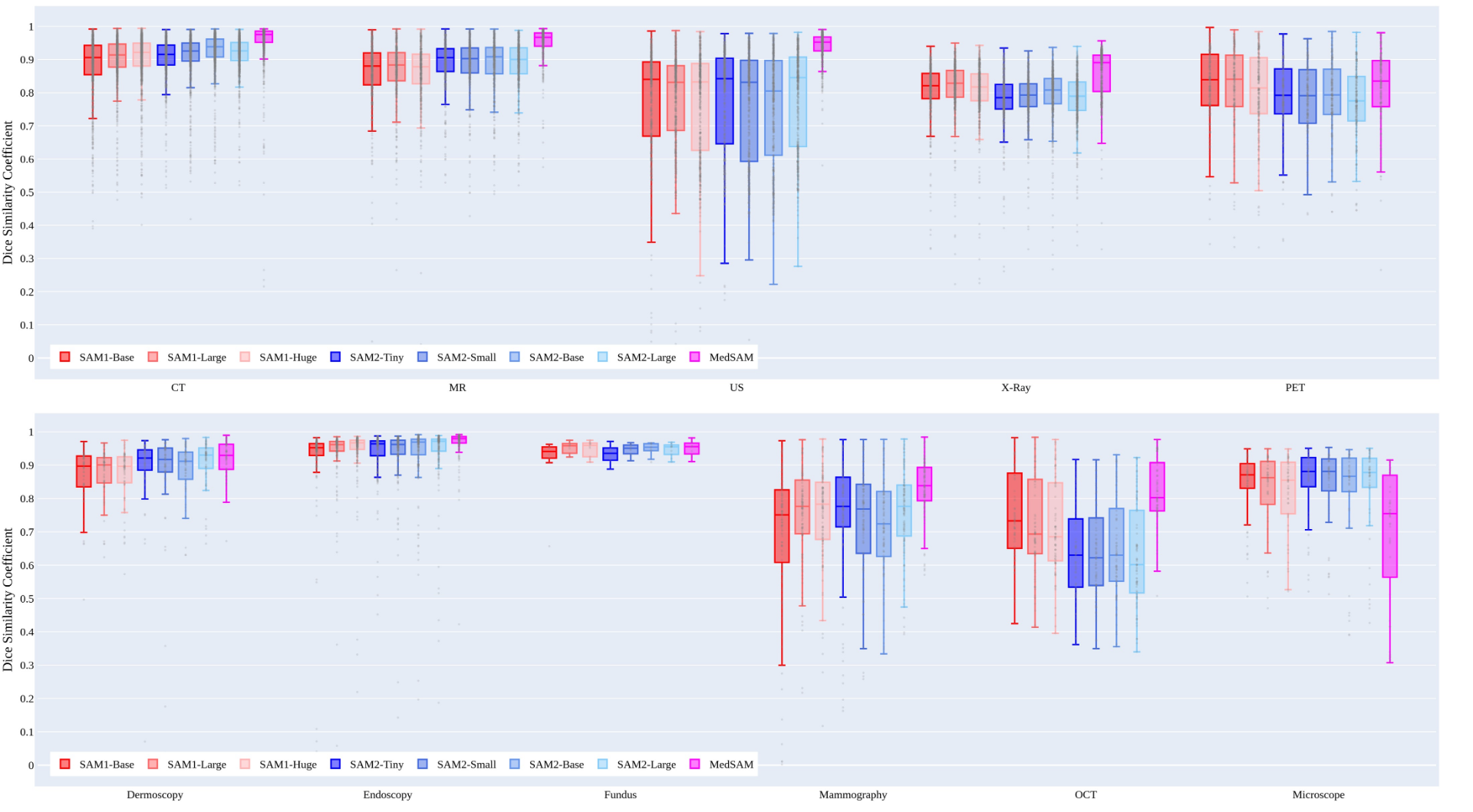}
\caption{Dot and box plot of the DSC scores for 2D image segmentation on 11 modalities. The plot shows descriptive statistics with the median value represented by the horizontal solid line within the box, the lower and upper quartiles delineating the borders of the box and the vertical black lines indicating 1.5$\times$IQR.}
\label{fig:2D-results}
\end{figure}

The segmentation results of 2D images across 11 modalities are presented in Fig.~\ref{fig:2D-results} and Table~\ref{tab:2D-DSC}. 
SAM2 obtained higher DSC scores than SAM1 in MR, dermoscopy, and light microscopy images but lower scores in PET, and OCT images.  
Their performance is comparable in CT, X-Ray, ultrasound, endoscopy, fundus, and mammography images. 
In contrast, MedSAM consistently outperformed both SAM1 and SAM2 in nine of the 11 modalities, except the PET and light microscopy images because its training set did not contain these types of images.

\begin{table}[htbp]
\caption{Quantitative segmentation results (average DSC scores) of eight SAM model variants on 2D images across 11 modalities. SAM1-Base: 93.7M; SAM1-Large: 312.3M; SAM1-Huge: 641.1M; SAM2-Tiny: 38.9M; SAM2-Small: 46.0M; SAM2-Base: 80.8M; SAM2-Large: 224.4M; MedSAM: 93.7M.}\label{tab:2D-DSC}
\centering
\begin{adjustbox}{width=0.99\textwidth}
\begin{tabular}{cccc|cccc|c}
\hline
Modality    & \multicolumn{1}{c}{SAM1-B} & \multicolumn{1}{c}{SAM1-L} & \multicolumn{1}{c|}{SAM1-H} & \multicolumn{1}{c}{SAM2-T} & \multicolumn{1}{c}{SAM2-S} & \multicolumn{1}{c}{SAM2-B} & \multicolumn{1}{c|}{SAM2-L} & MedSAM \\ \hline
CT               & 0.8825          & 0.9001 & 0.9028 & 0.9058          & 0.9090 & 0.9242 & 0.9167 & \textbf{0.9572} \\
MR               & 0.8620          & 0.8673 & 0.8602 & 0.8897          & 0.8867 & 0.8858 & 0.8863 & \textbf{0.9507} \\
PET              & \textbf{0.8198} & 0.8151 & 0.8042 & 0.7874          & 0.7791 & 0.7877 & 0.7727 & 0.8160          \\
Ultrasound       & 0.7749          & 0.7789 & 0.7703 & 0.7792          & 0.7614 & 0.7540 & 0.7873 & \textbf{0.9398} \\
X-Ray            & 0.8095          & 0.8181 & 0.8065 & 0.7835          & 0.7851 & 0.7939 & 0.7822 & \textbf{0.8573} \\
Dermoscopy       & 0.8683          & 0.8731 & 0.8706 & 0.8927          & 0.8706 & 0.8875 & 0.9068 & \textbf{0.9197} \\
Endoscopy        & 0.9170          & 0.9328 & 0.9395 & 0.9370          & 0.9346 & 0.9338 & 0.9315 & \textbf{0.9673} \\
Fundus           & 0.9119          & 0.9522 & 0.9495 & 0.9338          & 0.9475 & 0.9483 & 0.9465 & \textbf{0.9498} \\
Mammography      & 0.6931          & 0.7485 & 0.7512 & 0.7495          & 0.7304 & 0.7197 & 0.7601 & \textbf{0.8320} \\
OCT              & 0.7482          & 0.7148 & 0.7160 & 0.6414          & 0.6374 & 0.6498 & 0.6273 & \textbf{0.8166} \\
Light Microscope & 0.8332          & 0.8246 & 0.8178 & \textbf{0.8431} & 0.8388 & 0.8188 & 0.8322 & 0.6873          \\ \hline

\end{tabular}
\end{adjustbox}
\end{table}

\subsection{Evaluation results on 3D images}
\begin{figure}[h]
\centering
\includegraphics[scale=0.18]{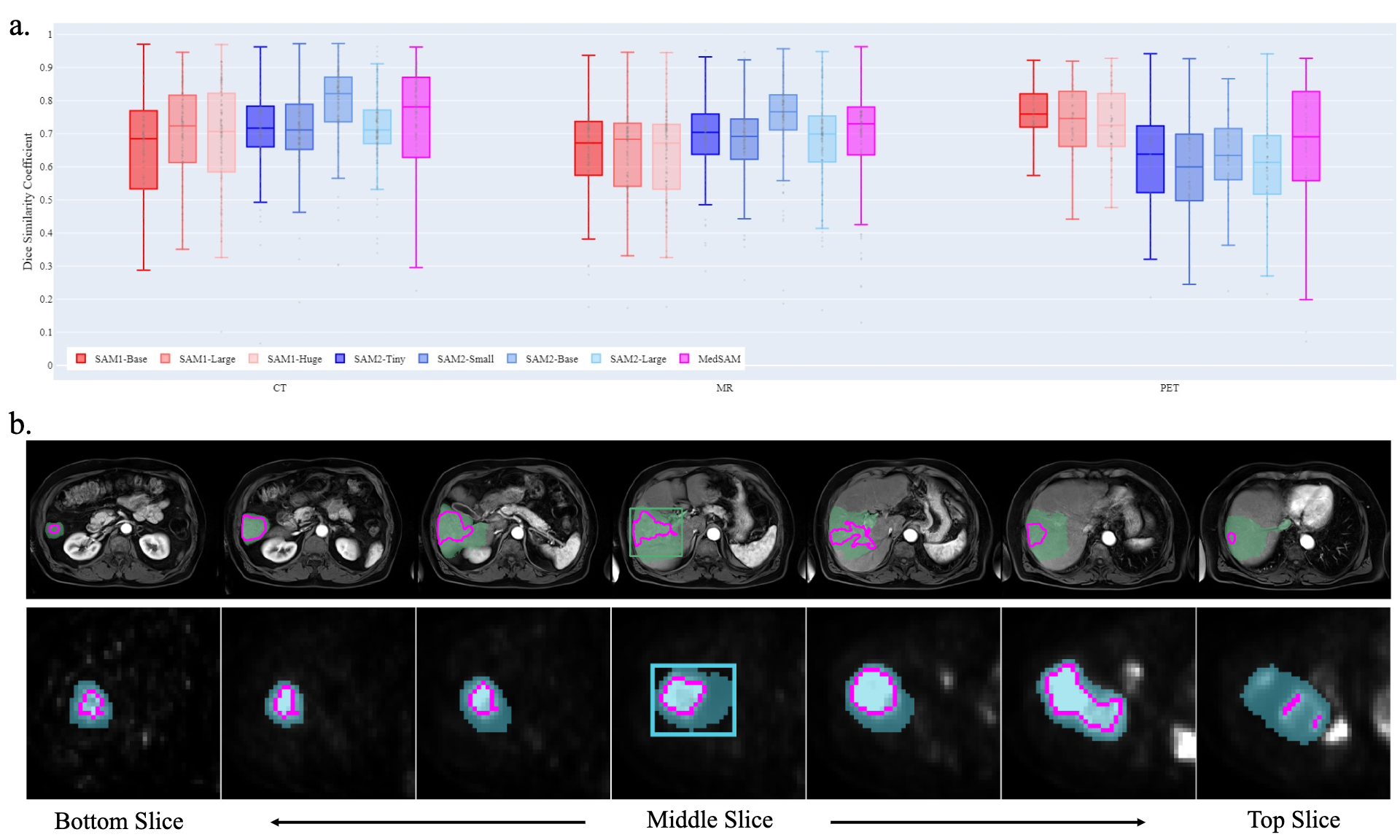}
\caption{\textbf{a}, Dot and box plot of the DSC scores for 3D image segmentation. The plot shows descriptive statistics with the median value represented by the horizontal solid line within the box, the lower and upper quartiles delineating the borders of the box and the vertical black lines indicating 1.5$\times$IQR. \textbf{b}, Visualized MR and PET segmentation examples of the best-performing SAM2. The bounding box prompt is initialized on the middle slicer and the generated mask is propagated to the top and bottom slices, respectively.}
\label{fig:3D-results}
\end{figure}

One of the major improvements of SAM2 is the video segmentation capability. Fig.~\ref{fig:3D-results} and Table~\ref{tab:3D-results} shows the quantitative results of SAM1, SAM2, and MedSAM in 3D CT, MRI, and PET images.
By formulating the 3D images as videos and propagating the mask from the middle slice to the remaining slices, SAM2-Base model achieved remarkable improvements over SAM1 and MedSAM on CT and MR images that processed the 3D data slice-by-slice. However, all SAM1 models outperformed SAM2 in PET images because the model tends to generate over-segmentaion errors in the middle slice and the error will propagate to the remaining slices as shown in Fig.~\ref{fig:3D-results}b.

\begin{table}[htbp]
\caption{Quantitative segmentation results on 3D images.}\label{tab:3D-results}
\centering
\begin{tabular}{ccccccc}
\hline
\multirow{2}{*}{Models} & \multicolumn{2}{c}{CT} & \multicolumn{2}{c}{MR} & \multicolumn{2}{c}{PET} \\ \cline{2-7} 
                        & DSC        & NSD       & DSC        & NSD       & DSC        & NSD        \\ \hline
SAM1-B                  & 0.6557          & 0.6492          & 0.6433          & 0.6575          & \textbf{0.7608} & \textbf{0.6393} \\
SAM1-L                  & 0.6982          & 0.6990          & 0.6486          & 0.6639          & 0.7411          & 0.6137          \\
SAM1-H                  & 0.6862          & 0.6848          & 0.6359          & 0.6496          & 0.7280          & 0.5877          \\ \hline
SAM2-T                  & 0.7168          & 0.6729          & 0.6919          & 0.6813          & 0.6305          & 0.3937          \\
SAM2-S                  & 0.7134          & 0.6679          & 0.6868          & 0.6733          & 0.5932          & 0.3444          \\
SAM2-B                  & 0.7199          & 0.6812          & 0.6758          & 0.6579          & 0.6331          & 0.3971          \\
SAM2-L                  & 0.7178          & 0.6795          & 0.6820          & 0.6673          & 0.6009          & 0.3400          \\ \hline
MedSAM                  & \textbf{0.7376} & \textbf{0.7434} & \textbf{0.6998} & \textbf{0.7172} & 0.6600          & 0.5788          \\ \hline
\end{tabular}
\end{table}

\subsection{Effects of mask initialization on 3D segmentation performance}
We further analyzed how the different 2D mask initializations of the middle slice affect the segmentation performance during the mask propagation in SAM2. In particular, we initialized the middle slice 2D mask with MedSAM-generated segmentation and the ground-truth mask, respectively. Table~\ref{tab:initialization} shows the quantitative segmentation in the three 3D modalities. Compared to the default settings that used SAM2 to segment the middle slice, using MedSAM-generated mask improved the final 3D segmentation across all three modalities with up to 17.5\% and 33.3\% in DSC and NSD scores, respectively. Furthermore, using the ground truth of the middle slice as the initialization can bring larger accuracy gains across the modalities and model sizes.

\begin{table}[htbp]
\caption{Quantitative segmentation results on 3D images with three different initializations for the middle slice.  Default: Using the corresponding SAM2 to generate the mask. MedSAM: Using MedSAM to generate the mask. GT: Using the ground-truth mask of the middle slice.}\label{tab:initialization}
\centering
\begin{tabular}{lcccccc}
\hline
\multicolumn{1}{c}{\multirow{2}{*}{\begin{tabular}[c]{@{}c@{}}Models \\ (Initialization)\end{tabular}}} & \multicolumn{2}{c}{CT} & \multicolumn{2}{c}{MR} & \multicolumn{2}{c}{PET}                           \\ \cline{2-7} 
\multicolumn{1}{c}{}                                                                                    & DSC        & NSD       & DSC        & NSD       & \multicolumn{1}{c}{DSC} & \multicolumn{1}{c}{NSD} \\ \hline
SAM2-T (Default)                                                                                        & 0.7168          & 0.6729          & 0.6919          & 0.6813          & 0.6305          & 0.3937          \\
SAM2-T (MedSAM)                                                                                         & 0.7790          & 0.7840          & 0.7600          & 0.7779          & 0.7661          & 0.6826          \\
SAM2-T (GT)                                                                                             & 0.8388          & 0.8386          & \textbf{0.8158} & \textbf{0.8331} & \textbf{0.8257} & \textbf{0.7730} \\ \hline
SAM2-S (Default)                                                                                        & 0.7134          & 0.6679          & 0.6868          & 0.6733          & 0.5932          & 0.3444          \\
SAM2-S (MedSAM)                                                                                         & 0.7830          & 0.7883          & 0.7537          & 0.7685          & 0.7682          & 0.6778          \\
SAM2-S (GT)                                                                                             & 0.8431          & 0.8443          & 0.8104          & 0.8245          & 0.8152          & 0.7570          \\ \hline
SAM2-B (Default)                                                                                        & 0.7199          & 0.6812          & 0.6758          & 0.6579          & 0.6331          & 0.3971          \\
SAM2-B (MedSAM)                                                                                         & 0.7875          & 0.7966          & 0.7487          & 0.7630          & 0.7656          & 0.6672          \\
SAM2-B (GT)                                                                                             & 0.8419          & 0.8453          & 0.8027          & 0.8039          & 0.8159          & 0.7551          \\ \hline
SAM2-L (Default)                                                                                        & 0.7178          & 0.6795          & 0.6820          & 0.6673          & 0.6009          & 0.3400          \\
SAM2-L (MedSAM)                                                                                         & 0.7881          & 0.7991          & 0.7552          & 0.7754          & 0.7594          & 0.6642          \\
SAM2-L (GT)                                                                                             & \textbf{0.8519} & \textbf{0.8634} & 0.8107          & 0.8263          & 0.8081          & 0.7511          \\ \hline
\end{tabular}
\end{table}

\subsection{Evaluation results on video segmentation}
\begin{figure}[htbp]
\centering
\includegraphics[scale=0.17]{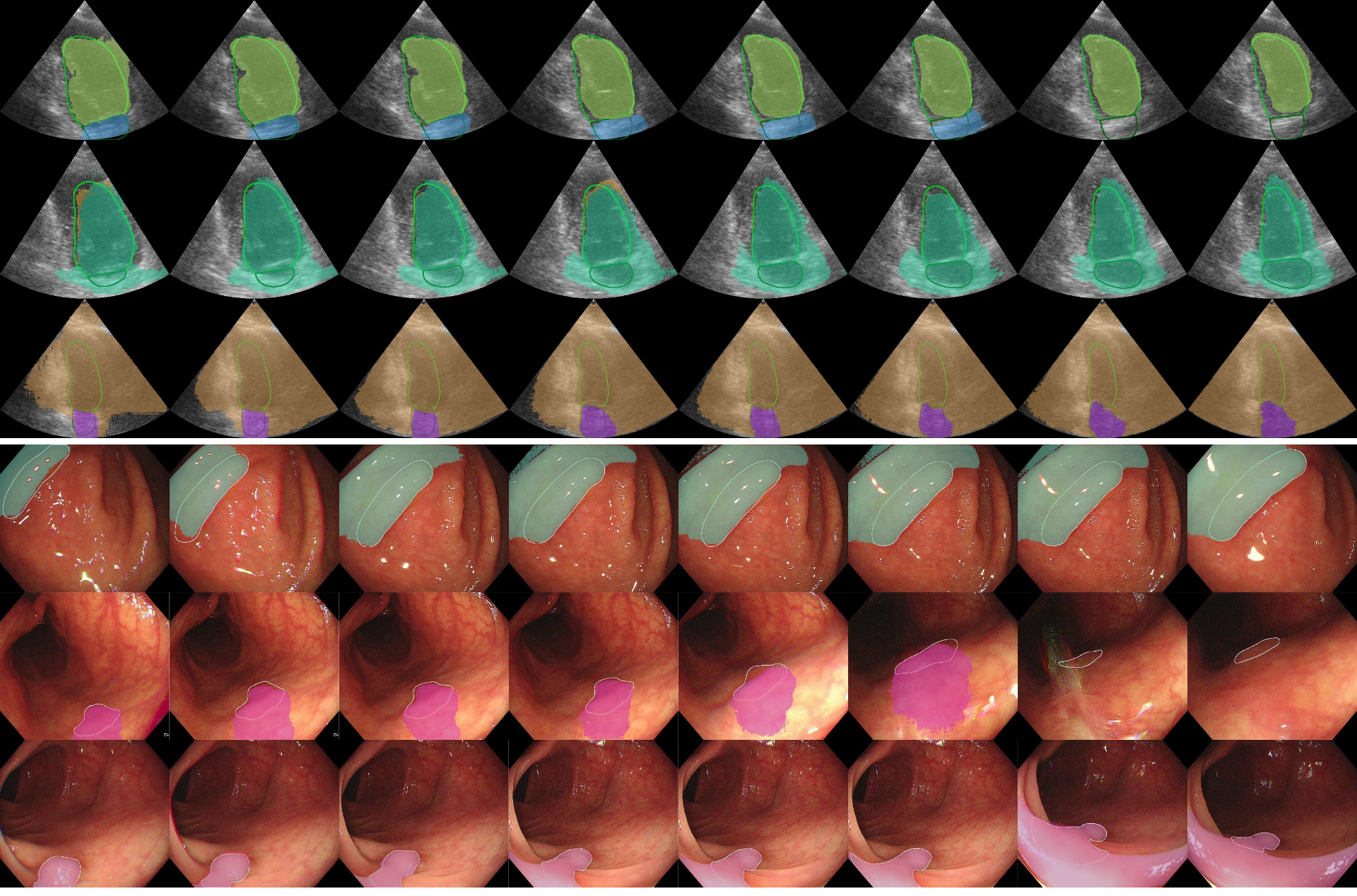}
\caption{Visualized examples of the best video segmentation model. The model either failed to segment the first frame or generated over-segmentation errors during the mask propagation when the object boundary is not clear or the images have low contrast.}
\label{fig:video-results}
\end{figure}

Next, we directly applied SAM2 for video segmentation where the first frame was initialized with box prompts. 
Table~\ref{tab:video} shows the heart and polyp segmentation results of SAM2 in echocardiography~\cite{Video-CAMUS} and endoscopy~\cite{Video-SUN-Data,Video-SUN-Seg-Gepeng} videos, respectively. SAM2-T achieved the best performance with a DSC score of 0.8537 for ultrasound videos while SAM-B obtained the highest DSC score of 0.8397 for endoscopy videos. We show some failure cases in Fig.~\ref{fig:video-results} of the best-performing model and it can be found that the model failed to segment the first frame or generate over-segmentation errors during the mask propagation when the object boundary is not clear or the images have low contrast.

\begin{table}[htbp]
\caption{Quantitative segmentation results on ultrasound and endoscopy video datasets.}\label{tab:video}
\centering
\begin{tabular}{ccccccccc}
\hline
\multirow{2}{*}{Modality} & \multicolumn{2}{c}{SAM2-T} & \multicolumn{2}{c}{SAM2-S} & \multicolumn{2}{c}{SAM2-B} & \multicolumn{2}{c}{SAM2-L} \\ \cline{2-9} 
                       & DSC          & NSD         & DSC          & NSD         & DSC          & NSD         & DSC          & NSD         \\ \hline
Ultrasound                & \multicolumn{1}{r}{\textbf{0.8537}} & \textbf{0.8824} & \multicolumn{1}{r}{0.7457} & 0.7602 & \multicolumn{1}{r}{0.8365}          & 0.8654          & \multicolumn{1}{r}{0.8358} & 0.8659 \\
Endoscopy                 & \multicolumn{1}{r}{0.8263}          & 0.8385          & \multicolumn{1}{r}{0.7954} & 0.8046 & \multicolumn{1}{r}{\textbf{0.8397}} & \textbf{0.8506} & \multicolumn{1}{r}{0.8121} & 0.8230 \\ \hline
\end{tabular}
\end{table}

\subsection{Transfer learning: a case study on adapting SAM2 to the medical domain}
Although SAM2 brings multiple improvements on top of SAM1, there is still a need to enhance its segmentation capability in medical image segmentation. Transfer learning has been demonstrated as an effective way to adapt SAM to new domain~\cite{MedSAM,SAM1-HQ}. However, SAM2 team has no plan to release the training code. Thus, we developed a transfer learning pipeline to fine-tune SAM2 in medical images.  

We used SAM2-T model as an example and conducted experiments on the 3D abdominal organ segmentation dataset~\cite{FLARE22-LDH} (Methods). The model was trained with 2D CT slices but the inference was conducted in 3D with initialization only on the middle slice, followed by a slice-wise propagation with the trained models.  
Table~\ref{tab:fine-tune} shows the comparison between SAM2-T and the fine-tuned model. It can be found that the fine-tuned model achieved remarkable improvements of $3.5-45.62$\% and $10.1-61.4$\% for DSC score and NSD scores, respectively.

\begin{table}[htbp]
\caption{Quantitative segmentation results on transfer learning with SAM-T for abdominal 3D organ segmentation in CT scans.}\label{tab:fine-tune}
\centering
\begin{tabular}{lcccccc}
\hline
\multicolumn{1}{c}{\multirow{2}{*}{Organ}} & \multicolumn{2}{c}{SAM2} & \multicolumn{2}{c}{SAM2 Fine-tuning} & \multicolumn{2}{c}{Improvements} \\ \cline{2-7} 
\multicolumn{1}{c}{}                       & DSC         & NSD        & DSC               & NSD              & DSC             & NSD            \\ \hline
Liver                                      & 0.5802      & 0.3605     & 0.9681            & 0.9127           & 38.79\%         & 55.22\%        \\
Right Kidney                               & 0.9059      & 0.8098     & 0.9410            & 0.9108           & 3.51\%          & 10.10\%        \\
Spleen                                     & 0.8040      & 0.6505     & 0.9601            & 0.9584           & 15.61\%         & 30.79\%        \\
Pancreas                                   & 0.1682      & 0.1138     & 0.4901            & 0.4732           & 32.19\%         & 35.95\%        \\
Aorta                                      & 0.1835      & 0.1338     & 0.6397            & 0.6512           & 45.62\%         & 51.73\%        \\
Inferior Vena Cava                         & 0.1438      & 0.1132     & 0.3468            & 0.3961           & 20.31\%         & 28.30\%        \\
Right Adrenal Gland                        & 0.3649      & 0.4126     & 0.6509            & 0.8009           & 28.60\%         & 38.84\%        \\
Left Adrenal Gland                         & 0.3195      & 0.3607     & 0.7095            & 0.8118           & 39.00\%         & 45.11\%        \\
Gallbladder                                & 0.6656      & 0.5412     & 0.8058            & 0.8481           & 14.02\%         & 30.70\%        \\
Esophagus                                  & 0.3342      & 0.2614     & 0.7951            & 0.8718           & 46.09\%         & 61.04\%        \\
Stomach                                    & 0.4333      & 0.2555     & 0.8452            & 0.7742           & 41.20\%         & 51.88\%        \\
Left Kidney                                & 0.8694      & 0.7700     & 0.9589            & 0.9247           & 8.95\%          & 15.47\%        \\ \hline
\end{tabular}
\end{table}

\subsection{Deployment: 3D Slicer Plugin and Gradio API for effective medical data annotation}
Numerous medical image segmentation models have been published and new models are continuously increasing every month, but only few of them are adopted by medical professionals. One of the main barriers is the lack of a user-friendly interface for medical professionals to access the model (without any coding) and provide feedback. SAM2 has provided an online interface to test the model but only short videos are allowed to be uploaded, and it also does not support most of the 3D medical data format.

\begin{figure}[htbp]
\centering
\includegraphics[scale=0.18]{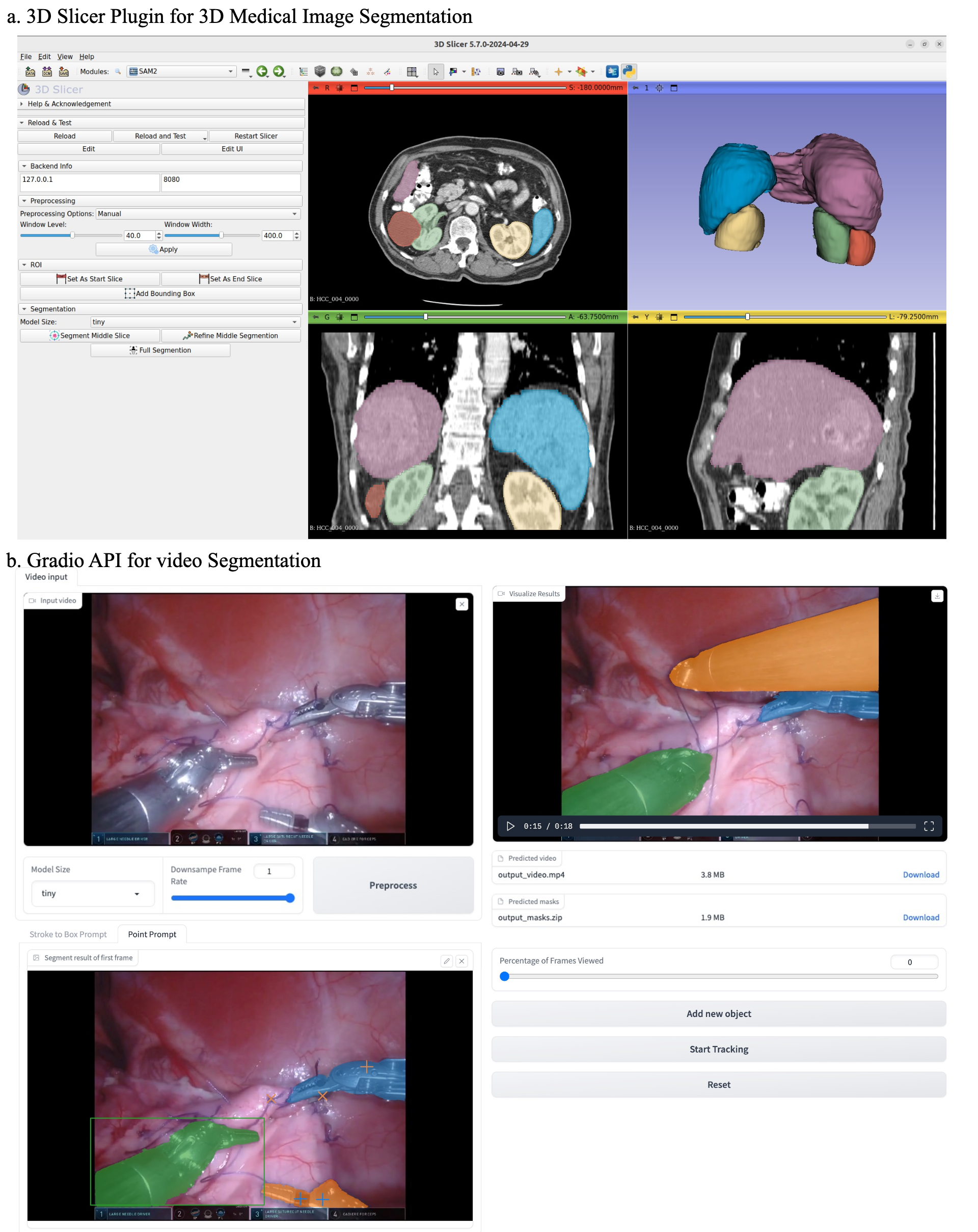}
\caption{SAM2 deployment for medical image and video annotation. \textbf{a}, 
Slicer plugin for 3D medical image segmentation. \textbf{b}, Gradio API for video segmentation.}
\label{fig:interface}
\end{figure}

In order to make SAM2 accessible to a wider audience, especially medical professionals, we developed two interfaces for general 3D medical image and video segmentation based on 3D Slicer~\cite{Slicer} and Gradio~\cite{Gradio}, respectively. Fig.~\ref{fig:interface} shows an overview of the two interfaces. 3D Slicer supports most of the common medical data formats and has built-in tools for user interactions, such as drawing bounding boxes, adding points, and revising masks.  
Our SAM2 3D Slicer plugin (Fig.~\ref{fig:interface}a) was designed for general 3D medical image segmentation with bounding box prompts. Users first specify the top and bottom slices of the segmentation target and draw a bounding box prompt in the middle slice. Then, SAM2 automatically generates the 2D mask on the middle slice and propagates it to the remaining slices. 
The Gradio API (Fig.~\ref{fig:interface}a) allows users to upload videos without time limitations. The initial frame can be any of the video frames and users can draw bounding boxes, point prompts, or their combinations to segment the target object. Then, SAM2 automatically tracks and segments the object across all the frames.  
We also make the API publicly available so that users can deploy it on their computing resources.

\section{Discussion}
\subsubsection*{Model Variants: SAM1 versus SAM2}
When SAM2 came out, a straightforward question emerged: "Does SAM2 outperform SAM1 in medical image segmentation?".
The answer could be "Yes" for 2D MRI, dermoscopy, and light microscopy images. For 3D CT and MRI scans, SAM2-B also achieved the highest scores, which outperformed all SAM1 models by a large margin. 
However, the answer could also be "No" for 2D OCT and PET images where SAM1 had better performance.  
One can attribute the improvements of SAM2 to the fact that it had advanced network architecture and was trained on both large-scale image and video datasets. The performance drop on some modalities could be due to the smaller model size compared to SAM1. Nevertheless, considering the sophisticated model development process, understanding the reasons for this performance gap requires access to the model training code and conducting rigorous ablation studies.

\subsubsection*{Scaling Laws: Model Size versus Performance}
Large models usually have a greater capacity for better performance, but we found that segmentation performance varies greatly across different model sizes on various medical image segmentation tasks. For example, the largest SAM2 model, SAM2-L, did not obtain the best performance in 8 of the 11 2D modalities and all the 3D modalities and videos. Notably, for 3D PET, light microscope images, and ultrasound videos, the smallest SAM-T model achieved the best performance. This indicates that model size alone is not a determinant of success in medical image segmentation tasks. Other factors, such as the specific characteristics of the training dataset and training protocols, likely play crucial roles in achieving optimal performance.

\subsubsection*{General-purpose Model (SAM2) versus Adapted Model in Medical Images (MedSAM)}
The video segmentation capability in SAM2 significantly enhances performance for 3D medical image segmentation, but it remains inferior to MedSAM for most 2D medical image modalities. The transfer learning experiments demonstrate that SAM2 can benefit from transfer learning to improve its ability to segment medical images. Nevertheless, it should be noted that direct fine-tuning can compromise its original versatile segmentation capabilities, as evidenced by the 2D image segmentation results. 
Therefore, it is important to consider a balance between leveraging transfer learning for improved task-specific performance and maintaining the general segmentation abilities of SAM2.

\subsubsection*{Limitation and Future Work}
The benchmark study has covered the common medical image modalities, but it can be further enhanced by including more 3D modalities, such as 3D ultrasound and OCT images. There are several open questions worth exploring further.
First, the video segmentation capability in SAM2 has broad applications in the medical domain, but the current model often fails to segment targets without clear boundaries. This issue can be addressed by transfer learning on medical datasets. Although fine-tuning video segmentation model is much more complicated, our 2D image transfer learning pipeline offers a good foundation for further development.
Second, SAM2 only supports point, box, and mask prompts. In contrast, text prompts offer greater flexibility for complex structure segmentation~\cite{BiomedParse,Yao-sam-text,EVF-SAM}. Implementing natural language processing capabilities within SAM2 could bridge the gap between complex medical terminology and model understanding, facilitating a more intuitive and efficient user experience.
Another critical direction is to make SAM2 more lightweight. SAM2-T is much smaller than SAM1-B but still requires large GPU RAM for long video segmentation. It is necessary to further reduce the model size and improve the inference efficiency without compromising performance for deployment in a broader range of clinical settings.

In conclusion, we have conducted a comprehensive evaluation of SAM2 across various medical image segmentation modalities. The performance varies significantly on different model sizes and segmentation tasks, indicating new or larger models are always better in terms of segmentation accuracy. Moreover, we develop a fine-tuning pipeline and demonstrate its effectiveness in improving the performance for 3D medical image segmentation.
We also provide 3D Slicer plugins and Gradio API to facilitate the deployment of SAM2 for medical image and video segmentation. 
Although such engineering implementations are rarely presented in deep learing-based medical image segmentation papers, we believe they are important for the adoption in clinical practice. 
We have made all the code publicly available for further development.

\section*{Methods}
\subsubsection*{Data sources and preprocessing}
All the evaluated images and videos were collected from public datasets~\cite{TCIA,FLARE21-MIA,FLARE22-LDH,liver-tumorMR-miccai23,autoPET-data,autoPET-data2,MSD-Dataset,NeurIPS-CellSeg,Video-SUN-Data,Video-SUN-Seg-Gepeng,Video-CAMUS,FUNDUS-Fives,Xray-dental,data-Mammo,data-OCT-intraretinal}, which were used in the validation set in the CVPR 2024 Medical Image Segmentation on Laptop Challenge\footnote{https://www.codabench.org/competitions/1847/}. None of them was used in the MedSAM training set.
The CT images were preprocessed with intensity cutoff based on the typical window level and window width. MR and PET images were clipped to the 0.5 and 99.5 percentiles of the non-zero region intensities. Then the intensity values were normalized to $[0, 255]$ via min-max normalization. For the remaining modalities, the intensities are not changed. 
Finally, all the images are converted to npz format for batch inference.

\subsubsection*{Key differences between SAM1 and SAM2 in methodology}
SAM2~\cite{SAM2} is a natural extension of SAM1~\cite{SAM1} for video segmentation with a unified framework for image and video segmentation tasks. Technically, there are two main methodology improvements. First, the Vision Transformer (ViT)~\cite{ViT2020} is replaced with Hiera~\cite{Hiera}, which can extract multi-scale features to produce high-resolution segmentation details. For the interactive segmentation of videos, the memory attention module is used to condition the predictions of the current frame on the information up to the current time point. The transformer architecture takes the image and prompt features from the current frame as well as the memories of features and predictions from previous frames. Endowing the memory attention module with the memory bank, SAM2 effectively extends the SAM's prompt encoder, image encoder, and mask decoder model architecture to enable robust and consistent segmentation across video frames by leveraging temporal context.

\subsubsection*{Fine-tunig protocol}
We fine-tuned the pre-trained SAM2-Tiny model on the MICCAI FLARE22 abdomen CT training dataset~\cite{FLARE22-LDH} with a data splitting ratio of 80\% and 20\% for training and validation. During fine-tuning, the prompt encoder was frozen because it is domain agnostic, while the image encoder and mask decoder were updated, allowing the model to adapt to the specific characteristics of the CT scans.
The input images were resized to [1024, 1024] and normalized using z-score normalization. For such CT dataset, the preprocessed CT scan slice images were duplicated to create 3-channel images before inputting them into the model. The bounding box prompts were obtained from the ground truth masks at the scale of [1024, 1024], and random perturbation of bounding box coordinates was applied, with a maximum coordinate shift of 5 pixels to improve the model's robustness. The ground truth masks were resized to [256, 256] to match the output dimension of the mask decoder. 
We fine-tuned the model using an AdamW optimizer~\cite{ADAMW} with a learning rate of 6e-5 and a batch size of 16. The fine-tuning was conducted for 1000 epochs, with manual early stopping when the training loss plateaued.
We used an unweighted sum of Dice loss and cross-entropy loss as the loss function because this compound loss has been proven to be robust in various segmentation tasks~\cite{SegLossOdyssey}.

\subsubsection*{Evaluaiton metrics}
We follow the suggestions in Metrics Reloaded~\cite{MetricsReloaded} and employ Dice Similarity Coefficient (DSC) and Normalized Surface Distance (NSD) to evaluate the region and boundary overlap ratio, respectively. 

\subsubsection*{Acknowledgements} 
The authors of this paper highly appreciate Meta AI for making SAM2 publicly available to the community. 
We thank all the dataset owners for making the invaluable data publicly available. We also thank 3D Slicer and Gradio team for providing the user-friendly platforms. 

%
\bibliographystyle{splncs04}
\bibliography{ref}

\end{document}